\documentclass[superscriptaddress,showpacs]{revtex4}
\usepackage[dvips]{graphicx}
\usepackage{amsmath}
\usepackage{amssymb,amsmath}
\usepackage{dsfont}

\newcommand{\bs}[1]{\boldsymbol{#1}}
\def\ket#1{|#1\rangle}
\def\bra#1{\langle#1|}

\def\expect#1{\langle#1\rangle}

\def\re#1{\mathrm{Re}\left( #1 \right)}
\def\im#1{\mathrm{Im}\left( #1 \right)}

\begin{document}

\today

\title{Open system effects on slow light and electromagnetically induced
transparency}

\author{Jonas Tidstr\"om}
\affiliation{School of Information and Communication Technology,
Royal Institute of Technology (KTH), Electrum 229, SE-164 40 Kista,
Sweden}
\email{tidstrom@kth.se}

\author{Marie Ericsson}
\affiliation{Department of Quantum Chemistry, Box 518, SE-751 20 Uppsala,
Sweden}

\author{Erik Sj\"oqvist}
\affiliation{Department of Quantum Chemistry, Box 518, SE-751 20 Uppsala,
Sweden}
\affiliation{Centre for Quantum Technologies, National University of Singapore,
3 Science Drive 2, 117543 Singapore, Singapore}

\author{L. Mauritz Andersson}
\affiliation{School of Information and Communication Technology,
Royal Institute of Technology (KTH), Electrum 229, SE-164 40 Kista,
Sweden}

\begin{abstract}
The coherence properties of a three-level $\Lambda$-system influenced by a Markovian
environment are analyzed. A coherence vector formalism is used and a vector form of
the Lindblad equation is derived. Together with decay channels from the upper state,
open system channels acting on the subspace of the two lower states are investigated,
i.e., depolarization, dephasing, and amplitude damping channels. We derive an analytic
expression for the coherence vector and the concomitant optical susceptibility, and
analyze how the different channels influence the optical response. This response depends
non-trivially on the type of open system interaction present, and even gain can be obtained.
We also present a geometrical visualization of the coherence vector as an aid to understand
the system response.
\end{abstract}
\maketitle
\section{Introduction}
\label{sec:introduction}

A medium that is opaque to radiation on one of the transitions of a
$\Lambda$-system can be made
transparent by addressing a laser field on the second transition, a
phenomenon known as electromagnetically induced transparency (EIT)
\cite{fleischhauer2005eit, marangos1998eit}. Associated with this
transparency is an extremely small and dynamically controllable group
velocity \cite{hau1999lsr}. The pulse can be brought to a complete
stop and stored as an atomic coherence. Most impressive with respect
to storage time, is perhaps the storage of a light pulse in a solid
for a period greater than one second
\cite{longdell2005sls}. The process can be reversed, and the
light pulse released with the original pulse information intact
\cite{fleischhauer2000tps}, making the EIT mechanism a strong candidate to build a
quantum memory \cite{fleischhauer2002qmp}. Beside applications in
quantum information processing, the narrow spectral feature of EIT is
an important tool in spectroscopy \cite{krishna2005hrh}, high
precision magnetic sensors \cite{katsoprinakis2006gps}, and atomic
clocks \cite{knappe2004mac, vanier2005acb}. Furthermore, the slow light phenomena
can be used for optical delay lines and light storage, but there is a trade-off
between the delay of a pulse and the bandwidth
\cite{tidstrom2007dbp}. In order to increase the bandwidth, spatially
dispersed light beams in non-isotropic media can be used
\cite{dutton2006aao}.

The above effects depend crucially on the coherence properties
of the $\Lambda$-system \cite{arimondo1996cpt}.
The dark state superposition, central to the EIT mechanism, is fragile to external influences.
This motivates to study open system effects in the dynamics of this system.
Open system effects
suppress the dark state of the system and thus set the ultimate
limit on the storage time of light, the absorption, and the
spectral resolution of the transparency. There are tricks that one can
play in order to increase the efficiency of this coherence effect. In,
for example, Ref.~\cite{longdell2005sls} the effective decoherence
rate was decreased by a method called quantum bang-bang dynamic
control, inspired by refocusing techniques in nuclear magnetic
resonance~\cite{viola1998dsd}.

In this paper, we investigate various types of interactions between
the $\Lambda$-system and a Markovian environment modeled by the Lindblad-Kossakowski
equation~\cite{Lindblad1976ogq,gorini1976cpd}. Using the coherence vector
formalism~\cite{lendi1987emc,byrd2011gos} we find the asymptotic states ($t\rightarrow \infty$)
for the $\Lambda$-system and derive an analytical expression for the corresponding
optical response, i.e., the susceptibility of the medium, due to the dephasing,
depolarization, and three different amplitude damping channels acting on the two
lower states. Effects due to open system interactions are relevant in many experimental
instances and depends on the detailed structure of the system, the environment, and
the interaction; for possible configurations see, e.g.,~\cite{fleischhauer1992rer,Moeller2008}.
The approach used here for treating open system
effects is general and provides an option for visualization of the system response.

The outline of the paper is as follows. Section II contains a general theory for a
Markovian dynamics in a three-level system using coherence vector formalism. In
Section III, we apply this framework to a $\Lambda$-system undergoing open system
dynamics corresponding to dephasing, depolarization, and amplitude damping channels.
The results are presented in Section IV. We give a geometrical visualization of different
projections of the steady state solution to the coherence vector when different open systems
effects are applied. We derive an analytical expression for the susceptibility in the
presence of the open system effects. The paper ends with the conclusions.

\section{Open three-level system}
\label{sec:theory}
Let us begin by studying a general three-level quantum system and the
effects due to the interaction with a Markovian environment. In this case,
the density operator $\hat{\rho}$ satisfies the master equation
\begin{equation}
\label{eq:lindblad-master-equation}
\partial_t\hat{\rho} =
\frac{1}{i \hbar}[\hat{H},\hat{\rho}]+\mathcal{\hat L}(\hat{\rho}),
\end{equation}
where $\hat{H}$ is the Hamiltonian operator. The Liouville superoperator
$\hat{\mathcal{L}}$ is linear, describes the coupling with the
environment, and can be written on Lindblad form \cite{Lindblad1976ogq}
\begin{equation}
  \label{eq:Liouville-operator}
  \mathcal{\hat L}(\hat{\rho})
  =
  \sum_{\mathrm{k}}(
  \hat \gamma_{\mathrm{k}} \hat \rho \hat \gamma_{\mathrm{k}} ^{\dagger}
  - \tfrac{1}{2} \hat \gamma_{\mathrm{k}}^{\dagger} \hat \gamma_{\mathrm{k}}
\hat \rho
  - \tfrac{1}{2} \hat \rho \hat \gamma_{\mathrm{k}}^{\dagger} \hat
\gamma_{\mathrm{k}}  ),
\end{equation}
where $\hat \gamma_{\mathrm{k}}$ are Lindblad operators, and the subscript k denotes
the open system channels.

For a two-level system it is customary to use a coherence vector representation, called
a Bloch-vector representation. Similarly, for a three level system we can write
the density matrix $\rho$ and Hamiltonian matrix $H$ in a vector form as \cite{lendi2007nls}
\begin{eqnarray}
  \label{density-operator}
  \rho & = & \frac{1}{3}\mathds{1} + \, \bs x\cdot \bs\lambda,
  \nonumber \\
  \label{Hamiltonian-vector-notation}
  H & = & \hbar \omega_0 \mathds{1} + \hbar \bs\omega\cdot\bs\lambda .
\end{eqnarray}
Here, we have introduced the coherence vector $\bs{x} = x_i {\bf e}_i$ (repeated indices
summed from now on), the torque vector $\bs \omega = \omega_i \bs{e}_i$ associated with
the Hamiltonian and the SU(3) matrix generators  $\bs \lambda = \lambda_i \bs{e}_i$. The
basis  $\{ {\bf e}_i \}_{i=1}^{8}$ is real and orthonormal. We have used the standard
scalar product \mbox{` $\cdot$ '}, and $\mathds{1}$ is the $3\times 3$ identity matrix.
We use the traceless Gell-Mann matrices $\{\lambda_i\}_{i=1}^{8}$ as defined in Ref.~\cite{arvnid1997ngp}.
The length of the coherence vector is \mbox{$0 \leq |\bs x|^2 \leq 1/3$}, with
$|\bs x|^2=1/3$ for pure states and $|\bs x|^2=0$ for maximally mixed states. 

By inserting Eq. (\ref{density-operator}) into Eq. (\ref{eq:lindblad-master-equation})
we obtain
\begin{eqnarray}
\label{eq:masterequation-coherencevector}
\dot{\bs x}\cdot \bs\lambda & = & \frac{1}{i}
[\bs\omega\cdot \bs\lambda, \bs x \cdot \bs\lambda] +
\frac{1}{3}\mathcal{L}(\mathds{1}) +
\mathcal{L}(\mathbf{\bs x\cdot \bs\lambda}),
\end{eqnarray}
i.e., the dynamics reformulated in terms of Gell-Mann matrices.
The non-zero form of the Lindblad operators $\hat
\gamma_{\mathrm{k}}$ is written as $\hat{\gamma}_{\mathrm{k}} \doteq
g_0 \mathds{1} + \bs g_{\mathrm{k}}\cdot \bs\lambda$, where the symbol
`$\doteq$' stands for `represented by'. In this paper we will not
consider the contribution from $g_0$ as this term always can be interpreted as an extra contribution to the Hamiltonian~\cite{footnote_aat}. Thus, we focus on the `Lindblad
vectors' $\bs g_\mathrm{k}$.

In Appendix A, we obtain a $\bs \lambda$-independent form of
Eq.~(\ref{eq:masterequation-coherencevector}). As a result we obtain the complete
dynamics of the system, including open system effects, as an inhomogeneous
differential equation \cite{lidar2004qpt}
\begin{equation}
  \label{eq:11}
  \dot{ \bs x}
  =
  M \bs x + \bs b
\end{equation}
with the evolution matrix $M$ and vector ${\bs b}$ explicitly give in Appendix A.
For time-independent and invertible $M$,
the general solution of Eq. (\ref{eq:11}) takes the form
\cite{lendi1987emc,lendi2007nls}
\begin{equation}
  \label{eq:general_solution}
  \bs x
  =
  e^{Mt} \bs{x}_0
  -
  M^{-1} \bs{b},
\end{equation}
where $\bs{x}_0 = \bs{x} (0) + M^{-1} \bs{b}$, $\bs x(0)$ being the
initial coherence vector.

The evaluation of the exponential of $M$ is non-trivial,
since $M$ is not diagonalizable in general. To deal with this fact one
may resort to the Jordan normal form of $M$, i.e., $M=S J S^{-1}$,
where $J$ has non-zero entries only on the diagonal and the
super-diagonal, and $S$ is an invertible matrix. Expanding the
exponent in Eq.~(\ref{eq:general_solution}), we can write the general
solution as
\begin{equation}
  \label{eq:general_solution-Jordan}
  \bs x
  =
  Se^{Jt}S^{-1} \bs{x}_0
  -
  M^{-1} \bs{b}.
\end{equation}
The real part of the eigenvalues $\{ s_j \}_{j=1}^8$, $s_j
\in C$, of $M$ originates from
$\hat{\mathcal{L}}$ and must be zero or negative in order to preserve
non-negativity of the density operator.

We now look at the eigenvectors of $M$. Higher-dimensional Jordan Blocks
occur provided $M$ has linearly dependent eigenvectors corresponding
to degenerate eigenvalues. However, in all the cases we
investigate in this paper we only identify one-dimensional Jordan blocks.

We now ask what happens for different open system effects
corresponding to one-dimensional Jordan blocks. We focus on the case
where we have $s_{i\in I}=0$, $I$ being an index set, and
$\mathrm{Re}(s_{i\notin I})<0$. In the $t\rightarrow \infty$ limit,
the matrix $e^{Jt}$ tends to the matrix $P'$, which projects onto
the subspace corresponding to $I$. Thus, in this limit,
$e^{Mt}=Se^{Jt}S^{-1}$ tends to the projector $SP'S^{-1}$ and
the asymptotic coherence vector
${\bs x}^{(\infty)}$ depends on the initial coherence vector ${\bs x}_0$
according to
\begin{eqnarray}
\label{eq:asymptotic1}
{\bs x}^{(\infty)}=P {\bs x}_0-M^{-1}{\bs b} .
\end{eqnarray}
For the cases studied in this paper, $I$ is the empty set, i.e.,
$\mathrm{Re}(s_i)<0 \ \forall i$, we have $e^{Mt}=
Se^{Jt}S^{-1}\rightarrow 0$ when $t\rightarrow\infty$. In this
case, the asymptotic state is determined by
\begin{equation}
{\bs x}^{(\infty)}=-M^{-1} {\bs b} ,
\label{eq:assymptotic-general-solution}
\end{equation}
thus being independent of ${\bs x}_0$.
Further discussions on the existence and properties of asymptotic states 
in Markovian open quantum systems can be found in Refs.~\cite{Ticozzi2007,Ticozzi2009,Schirmer2010}.

\section{Open $\Lambda$-system}
\label{sec:g-m-b}
\begin{figure}[t]
  \centering
\includegraphics[width=4cm]{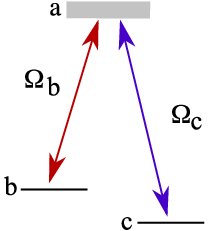}
\caption{The $\Lambda$-system. The field on the $a$-$b$ transition is the weak
probe field $\Omega_b$, and the field on the $a$-$c$ transition is the control
field $\Omega_c$. The upper level is broadened by the finite lifetime
$1/\gamma$.}
\label{fig:Lambdasystem}
\end{figure}
An electromagnetic field interacting with matter gives rise to a
matter polarization. The linear part of the polarization field
$P=P(E)$ can be written in terms of the electrical susceptibility
$\chi = P / (E \epsilon_0)$, with $\epsilon_0$ the electrical
permittivity in free space.
The real and imaginary parts of the susceptibility describe the phase
velocity of the electrical field and the absorption or gain of the
field, respectively. The macroscopic polarization field of an isotropic
medium consisting of $N$ atoms per unit volume is
$\mathbf{P}\equiv \expect{\hat{\mathbf{P}}}
= N \expect{\hat{\mathbf{p}}}
= N  \mathrm{Tr}[\hat \rho  \hat{\boldsymbol{\mu}} ],
$
where $\hat \rho$ is the density operator and $\hat{\boldsymbol{\mu}}
= e \,\hat{\mathbf{r}}$ is the dipole operator with electric charge
$e$ and position operator $\hat{\mathbf{r}}$. Therefore, if $\hat \rho$ is
known as a function of the optical field, the optical response is too,
as it is given by the coherences of $\hat \rho$.
In this section we develop the tools to compute $\hat \rho$ for an open $\Lambda$-system
and in the next section we find the resulting optical response $\chi$.

The $\Lambda$-system consists of two ground state levels
coupled to an excited state via electromagnetic fields, as shown
in Fig.~\ref{fig:Lambdasystem}. The corresponding Hamiltonian is
given by $\hat{H}(t) = \hat{H}_0-\hat{\boldsymbol\mu} \cdot
\mathbf{E}(t)$, where $\hat{H}_0$ is the Hamiltonian of the bare
atom, $\hat{\boldsymbol{\mu}}$ is the electric dipole operator, and
$\mathbf{E}(t)$ is an oscillating electric field. Furthermore,
let $\mathcal{E}_i$ be the energy eigenvalues of $\hat{H}_0$
corresponding to the bare eigenstates $\ket{i}$, $i=a,b,c$. In a frame
rotating with the external field, the Hamiltonian operator reads
\begin{equation}
  \label{eq:hamiltonian-opertor-form}
  \hat H
  =
  \hbar \, \delta_b \ket{b} \bra{b}
  +
  \hbar \, \delta_c \ket{c} \bra{c}
  -
  \hbar\,
  \Omega_b \ket{a} \bra{b}
  -
  \hbar\,
  \Omega_c \ket{a} \bra{c}
  +
  h.c.,
\end{equation}
where we have used the rotating wave approximation and we have defined
the complex Rabi frequencies in terms of the electric field $\mathbf{E}$
in the rotating frame as $\Omega_{i}=\bra{a} {\hat{\boldsymbol\mu}} \ket{i}
\cdot \mathbf{E}/ (2\hbar)$ and detunings $\delta_i = \omega_i -
(\mathcal{E}_a - \mathcal{E}_i)/\hbar$, where $\omega_i$ are the
frequencies of the field. Note the factor
1/2 in the definition of $|\Omega_i|$, introduced for
notational convenience. Defining the Gell-Mann matrices in terms
of the basis ordered as $\{ \ket{b},\ket{c},\ket{a}\}$, the
parameters of the Hamiltonian matrix in
Eq.~(\ref{Hamiltonian-vector-notation}) are given by
\begin{align}
  \label{eq:omega-omega0}
  \omega_0
  & =
  \frac{1}{3\hbar}
  \mathrm{Tr}[H]
  =
  \frac{2}{3}\Delta,
  \nonumber\\
  \bs \omega
  & =
  \frac{1}{2\hbar}
  \mathrm{Tr}[H  \lambda_i] \bs e_i
  =\delta {\bs e}_3 - |\Omega_b|\cos \phi_b {\bs e}_4
  -|\Omega_b|\sin \phi_b {\bs e}_5
    -|\Omega_c|\cos \phi_c{\bs e}_6
    -|\Omega_c|\sin \phi_c{\bs e}_7
-\frac{1}{ \sqrt 3}\Delta{\bs e}_8,
\end{align}
where $\Omega_i = |\Omega_i| \exp(i \phi_i) $ and we have defined
the two-photon detuning $\delta = (\delta_b-\delta_c)/2$ as well
as the mean detuning $\Delta = (\delta_b + \delta_c)/2$.

We now specify the Lindblad operators that describe the different
forms of open system dynamics. Spontaneous emission corresponding to the amplitude damping
of the excited energy state $\ket{a}$ to either of the two ground states
$\ket{b}$ or $\ket{c}$ with rates $\gamma_{b}$ and $\gamma_{c}$ is assumed to
be always present. The corresponding Lindblad operators and their matrix representation read
\begin{eqnarray}
  \label{eq:gb}
  \hat{\gamma}_{b}
  &=&
  \sqrt{\gamma_{b}} \ket{b} \bra{a}
  \doteq
  \frac{1}{2} \sqrt{\gamma_{b}}
  (\lambda_4 + i\lambda_5),
  \nonumber
  \\
  \label{eq:gc}
  \hat{\gamma}_{c}&=&
  \sqrt{\gamma_{c}}\ket{c} \bra{a}
  \doteq
  \frac{1}{2} \sqrt{\gamma_{c}}
  (\lambda_6 + i\lambda_7),
\end{eqnarray}
and they define the natural line width $\gamma \equiv \gamma_{b} +
\gamma_{c}$ indicated in Fig.~\ref{fig:Lambdasystem}.  The subscripts $b$ and $c$ indicate the final quantum state of the atom after interaction with the environment through the decay channel. The corresponding Lindblad vectors are
\begin{eqnarray}
  \label{eq:3332}
  \bs g_{b} & =& \frac{1}{2} \sqrt{\gamma_{b}} (\bs{e}_4+i\bs{e}_5),
  \nonumber
  \\
  \bs g_{c} & = &\frac{1}{2} \sqrt{\gamma_{c}} (\bs{e}_6+i\bs{e}_7).
\end{eqnarray}
These vectors are complex-valued and therefore $C^{(-)}$ in Eq.~(\ref{eq:C-matrix}) is non-vanishing and, accordingly, the decay channel contributes to the dynamics, Eq.~(\ref{eq:11}), with a non-zero ${\bs{b}}$.

The dark state of the $\Lambda$-system is
\begin{equation}
  \label{eq:dark-state}
  \ket{d}=\frac{1}{\sqrt{|\Omega_b|^2+|\Omega_c|^2}}
  (\Omega_c^* \ket{b} - \Omega_b^* \ket{c}).
\end{equation}
This is an eigenstate of the Hamiltonian, which is unaffected by the
decay channels defined by Eq.~(\ref{eq:gb}). The additional open
system effects that we are primarily interested in are the ones that
act on the two-dimensional subspace $V_{bc} = \{ \ket{b} , \ket{c}
\}$, since these could potentially destroy the dark state and the
associated phenomena. We consider the limit of a weak
probe field and a strong control field, i.e., $|\Omega_b| \ll
|\Omega_c|$, hence the dark state will be close to $\ket{b}$.

Although a continuum set of channels is necessary to fully cover
all possible cases \cite{Bacon2001,Lloyd2001}, it is known \cite{nielsen2000qci} that
depolarization, dephasing, and amplitude damping 
constitute a set of channels that captures essential features of 
open system effects for two-level systems. 
Here, we examine these channels acting on
$V_{bc}$. Depolarization and dephasing are particular combinations of
the Hermitian Lindblad operators
\begin{eqnarray}
  \label{eq:gamma-x}
  \hat{\gamma}_{x} & = &
  \sqrt{\eta_x} ( \ket{b} \bra{c} + \ket{c} \bra{b} )
  \doteq
  \sqrt{\eta_x}
  \lambda_1,
  \nonumber
\\
  \label{eq:gamma-y}
  \hat{\gamma}_y & = &
  \sqrt{\eta_y} ( -i\ket{b} \bra{c} + i \ket{c} \bra{b} )
  \doteq
  \sqrt{\eta_y}\lambda_2,
  \nonumber
\\
  \label{eq:gamma-z}
  \hat{\gamma}_z & = &
  \sqrt{\eta_z} ( \ket{b} \bra{b} - \ket{c} \bra{c} )
  \doteq
  \sqrt{\eta_z}
\lambda_3 ,
\end{eqnarray}
where the symbols $\eta_k$ are used to denote rates of the various
channels acting on $V_{bc}$. The corresponding
Lindblad vectors are real and given by
\begin{eqnarray}
  \label{eq:polar-decoherence-vector}
  \bs g_x & = \sqrt{\eta_x} \bs{e}_1,
  \nonumber
  \\
  \bs g_y & = \sqrt{\eta_y} \bs{e}_2,
  \nonumber
  \\
  \bs g_z & = \sqrt{\eta_z} \bs{e}_3.
\end{eqnarray}
The depolarization channel is isotropic, i.e., corresponds to
$\eta_x=\eta_y=\eta_z\equiv\eta/3$. The dephasing channel, on the
other hand, is anisotropic and corresponds to $\eta_x=\eta_y=0,
\eta_z\neq 0$. Neither depolarization nor dephasing contribute to
${\bs{b}}$ since $C^{(-)}$ vanishes.

As already stated, the dark state is close to $\ket{b}$ when the intensity of the
probe field is much smaller than the intensity of the control field. 
Because of this we expect a non-vanishing amplitude damping $b\leftarrow c$ 
to influence the $\Lambda$-system in a different manner as compared to the 
channel $c\leftarrow b$. To examine this quantitatively, we study the following channels
\begin{eqnarray}
  \label{eq:g_damp-bc}
  \hat{\gamma}_{bc}
  =
  \sqrt{\eta_{bc}} \ket{b} \bra{c}
  \doteq
  \frac{1}{2} \sqrt{\eta_{bc}}
  (\lambda_1 + i\lambda_2),
  \nonumber
  \\
  \label{eq:g_damp-cb}
  \hat{\gamma}_{cb}
  =
  \sqrt{\eta_{cb}} \ket{c} \bra{b}
  \doteq
  \frac{1}{2} \sqrt{\eta_{cb}}
  (\lambda_1 - i\lambda_2),
\end{eqnarray}
where $\hat{\gamma}_{bc}$ flips $\ket{c}$ to $\ket{b}$ at a rate
$\eta_{cb}$ and vice versa for $\hat{\gamma}_{cb}$. The corresponding
Lindblad vectors read
\begin{eqnarray}
  \label{eq:g-damp}
  \bs g_{bc} & = \frac{1}{2} \sqrt{\eta_{bc}} (\bs{e}_1+i\bs{e}_2),
  \nonumber
  \\
  \bs g_{cb} & = \frac{1}{2} \sqrt{\eta_{cb}} (\bs{e}_1-i\bs{e}_2),
\end{eqnarray}
which are complex-valued and may therefore contribute to
${\bs{b}}$.

The evolution matrix $M$ and the vector $\bs b$ are obtained by summing the contributions from the above described open system channels. We may write the resulting $M$ in a simple form by applying the following similarity transformation
\begin{equation}
	\label{eq:similarity-transformation}
	M = R^{-1} M' R,
\end{equation}
where
\begin{align}
  \label{eq:transformation-op}
  R
  =
  \bs{e}_1^{\phantom{\mathrm{T}}}\! \bs{e}_1^\mathrm{T}+
  \bs{e}_2^{\phantom{\mathrm{T}}}\! \bs{e}_2^\mathrm{T}+
  \bs{e}_3^{\phantom{\mathrm{T}}}\! \bs{e}_3^\mathrm{T}+
  \bs{e}_5^{\phantom{\mathrm{T}}}\! \bs{e}_4^\mathrm{T}+
  \bs{e}_6^{\phantom{\mathrm{T}}}\! \bs{e}_5^\mathrm{T}+
  \bs{e}_7^{\phantom{\mathrm{T}}}\! \bs{e}_6^\mathrm{T}+
  \bs{e}_8^{\phantom{\mathrm{T}}}\! \bs{e}_7^\mathrm{T}+
  \bs{e}_4^{\phantom{\mathrm{T}}}\! \bs{e}_8^\mathrm{T} .
\end{align}
The transformed evolution matrix $M'$ takes the block structure form
\begin{align}
  \label{eq:M-block-matrix}
  M'
  =
  \begin{pmatrix}
    A & C \\
    -C^\mathrm{T} & B
  \end{pmatrix}
\end{align}
with the submatrices
\begin{eqnarray}
  \label{eq:A}
  A
  &=&
  \begin{pmatrix}
    -\mathrm{Re}(\Gamma)& -\mathrm{Im}(\Gamma)& 0& 0 \\
    \mathrm{Im}(\Gamma)& -\mathrm{Re}(\Gamma)& 0& 0 \\
    0& 0& -\eta_{{}_{+}}& \tfrac{\eta_{{}_{-}}\!-\gamma_\mathrm{m}}{\sqrt{3}} \\
    0& 0& 0& -2\gamma \\
  \end{pmatrix} ,
  \nonumber\\
  \label{eq:B}
  B
  &=&
  \begin{pmatrix}
    -\mathrm{Re}(\Gamma_{-}) & -\mathrm{Im}(\Gamma_{-}) & 0 & 0 \\
    \mathrm{Im}(\Gamma_{-}) & -\mathrm{Re}(\Gamma_{-}) & 0 & 0 \\
    0& 0& -\mathrm{Re}(\Gamma_{+}) & -\mathrm{Im}(\Gamma_{+}) \\
    0& 0& \mathrm{Im}(\Gamma_{+}) & -\mathrm{Re}(\Gamma_{+}) \\
  \end{pmatrix} ,
  \nonumber\\
  \label{eq:C}
  C
  &=&
  \begin{pmatrix}
  |\Omega_c|\sin \phi_c & -|\Omega_c|\cos \phi_c &
  |\Omega_b|\sin \phi_b & -|\Omega_b|\cos \phi_b \\
  |\Omega_c|\cos \phi_c & |\Omega_c|\sin \phi_c &
  -|\Omega_b|\cos \phi_b & -|\Omega_b|\sin \phi_b \\
  |\Omega_b|\sin \phi_b & -|\Omega_b|\cos \phi_b &
  -|\Omega_c|\sin \phi_c & |\Omega_c|\cos \phi_c \\
  \sqrt{3}|\Omega_b|\sin \phi_b & -\sqrt{3}|\Omega_b|\cos \phi_b &
  \sqrt{3} |\Omega_c|\sin \phi_c & -\sqrt{3}|\Omega_c|\cos \phi_c
  \end{pmatrix} .
\end{eqnarray}
Here, we have introduced the parameters
\begin{eqnarray}
  \label{eq:parameters}
  \gamma_\mathrm{m} & = & \gamma_{b} - \gamma_{c},
  \nonumber\\
  \eta_{{}_{\pm}}  & = & \eta_{bc} \pm \eta_{cb},
  \nonumber\\
  \Gamma  &=&  2i \delta + (\eta_{{}_{+}} + 8 \eta + 4 \eta_z)/2,,
  \nonumber\\
  \Gamma_{\pm}  &=& i(\delta \mp \Delta) +  \gamma_{\pm},
  \nonumber\\
  \gamma_{\pm} & = & \gamma + (6 \eta +2 \eta_{z} +\eta_{{}_{+}}\pm \eta_{{}_{-}})/4.
\end{eqnarray}

The diagonal elements of $M'$ are real and negative, and contain the open system rates.
The only off-diagonal element containing open system effects is $(\eta_{{}_{-}} - \gamma_\mathrm{m})/\sqrt{3}$
in $A$. They are non-vanishing if the amplitude damping rates in $V_{bc}$, and also if the decay rate
from the exited state to the ground states, are unbalanced. This fact also influences $\bs{b}$, which
is explicitly given by
\begin{equation}
  \label{eq:explicit-b-vector}
  \bs b
  =
  \frac{2 \eta_{{}_{-}} + \gamma_\mathrm{m} }{6} \bs e_{3} +
  \frac{\gamma}{2 \sqrt{3}} \bs e_{8}.
\end{equation}
Thus, $\bs{b}$ is non-vanishing if damping processes are present in the open system dynamics.

\section{Results}

We begin by analyzing the asymptotic state of the quantum system as represented by the
coherence vector ${\bs x}^{(\infty)}$. The use of the asymptotic coherence vector is justified 
by its relation to measurable quantities. For instance, two of its components determine the real
and imaginary parts of the induced electric polarization according to $P_{ab} =
N \mu_{ab} \rho_{ab}^{(\infty)} = N \mu_{ab}(x_4^{(\infty)}+ix_5^{(\infty)})$, from which
the susceptibility of the probe field can be found.

\begin{figure}[th!]
  \centering
  \includegraphics[width=0.7\textwidth]{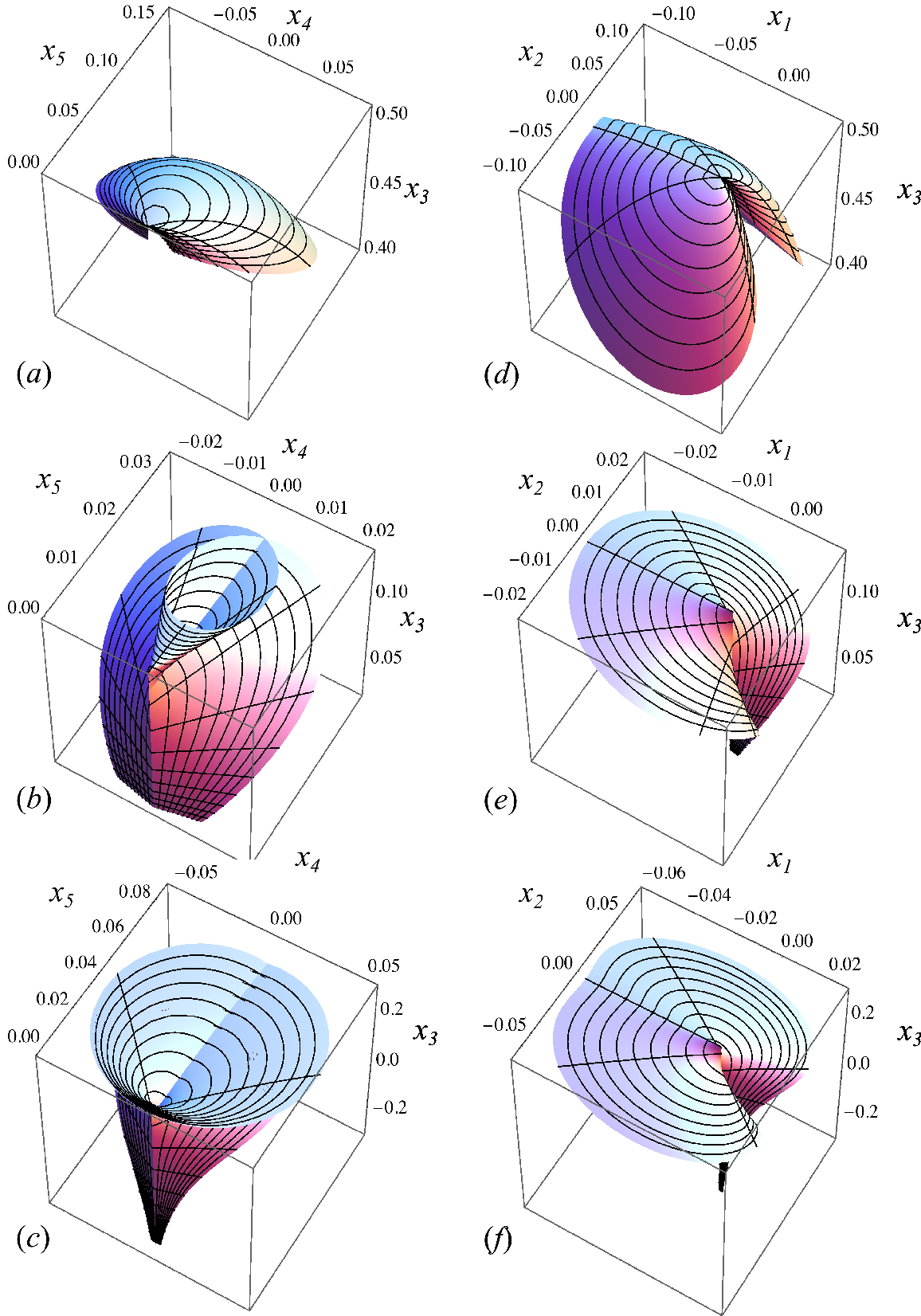}
  \caption{Plots of the asymptotic state for two different cuts through the eight dimensional coherence vector space is shown:  $x_4$, $x_5$, $x_3$ coordinates of the asymptotic state in (a-c) and $x_1$, $x_2$, $x_3$ coordinates of the asymptotic state in (d-f). Different open system channels are shown: ideal case in (a,d), depolarization $\eta=0.2$ in (b,e), and amplitude damping $c \leftarrow b$ channel $\eta_{cb}=0.2$ in (c,f). Radial mesh corresponds to changing $\Omega_b$ from \mbox{0 to 0.1} in steps of 0.02, tangential mesh corresponds to the two-photon detuning \mbox{$-5$ to 5} in steps of 0.5; the exception is (a) where the interval is given between \mbox{$-$5 to 0}. Other parameter values are $\Omega_c=1$, $\gamma=1$ and $\Delta=0$.}
  \label{fig:coherence-vector-visual}
\end{figure}

In Fig.~\ref{fig:coherence-vector-visual}, we show the projection of the eight-dimensional
coherence vector space onto three-dimensional cuts for the cases of depolarization and
amplitude damping $c \leftarrow b$ channels, and compare with the ideal case where
decay of the excited state $\ket{a}$ to the two ground states $\ket{b}$ and $\ket{c}$
are the only open system effects present. The two parameters spanning the surface are the
probe field strength $\Omega_b$ and the two photon detuning $\delta$; the control field
strength $\Omega_c$ is held constant.

The components $x_1, x_2$, and  $x_3$ shown in Fig.~\ref{fig:coherence-vector-visual}(d-f),
concern the coherence of the asymptotic state in the $V_{bc}$ subspace. Components 1 and 2
are the real and imaginary part of the coherence $\rho_{bc}$, i.e., they describe the phase
relation between the $\ket{b}$ and $\ket{c}$ amplitudes of the asymptotic state. The third
component $x_3$ is the corresponding population balance. Fig.~\ref{fig:coherence-vector-visual}(d)
shows the ideal case at zero detuning. Here, $x_1$ is negative, $x_2$ is zero, and $x_3$ is
close to 0.5. This is precisely the dark state with a real and negative coherence
between the $\ket{b}$ and $\ket{c}$ amplitudes, and most of the population in $\ket{b}$.
For non-zero detuning, we can visualize how the dark state deteriorates, by inspecting the
entire surface. Depolarization is added in Fig.~\ref{fig:coherence-vector-visual}(e). In this
case, we see how the $x_3$ component radically changes towards a more balanced population and
the magnitudes of the coherences $x_1$ and $x_2$ decrease, as compared to the ideal case.
This can be understood physically as the depolarization channel drives the system towards a 
balanced mixture in the $V_{bc}$ subspace. 
Similar effects occur for the case of the $c \leftarrow b$ channel, shown
in Fig.~\ref{fig:coherence-vector-visual}(f).

The components $x_3, x_4$, and $x_5$ shown in Fig.~\ref{fig:coherence-vector-visual}(a-c)
are directly related to the optical response of the atom to the probe field. A zero detuned
ideal case is characterized by vanishing $x_4$ and $x_5$. This implies that the electric 
polarization vanishes, clearly demonstrating the EIT effect. By adding depolarization,
the $x_5$ component becomes positive at zero detuning and therefore absorption is present, 
as shown in Fig.~\ref{fig:coherence-vector-visual}(b). The $c \leftarrow b$ amplitude damping
channel, shown in Fig.~\ref{fig:coherence-vector-visual}(c), gives rise to a negative $x_5$ 
component, corresponding to probe field gain.

Let us now examine the susceptibility for a small probe field that
interacts with a medium consisting of a collection of
$\Lambda$-systems that in turn are affected by the open system
channels considered in Sec.~\ref{sec:g-m-b}. The susceptibility \cite{boyd2003no}, of a medium is related
to the asymptotic state of the atomic coherence $\rho_{ab}^{(\infty)}
= x_4^{(\infty)}+ix_5^{(\infty)}$, which can be calculated by
Eq.~(\ref{eq:assymptotic-general-solution}) using Eqs.~(\ref{eq:M-block-matrix}) and 
(\ref{eq:explicit-b-vector}). Explicitly,
\begin{equation}
\chi = \lim_{\Omega_b\rightarrow 0} \kappa \frac{\partial}{\partial \Omega_b } \rho_{ab}^{(\infty)}
\label{eq:susceptiblity-def} 
\end{equation}
with $\kappa = N |\bra{a} {\hat{\boldsymbol\mu}} \ket{b}|^2/(2 \epsilon_0 \hbar)$, and we 
recall that the Rabi frequency is defined in a non-standard way with an extra factor of $2$ 
for notational convenience. The expression of the susceptibility in
Eq.~(\ref{eq:susceptiblity-def}) holds in the limit of a small probe field. Furthermore, 
we recall that the real part of the susceptibility is related to the phase evolution of the 
optical field, while the imaginary part describes the absorption or gain.

The open system channels defined by the Lindblad vectors in
Eqs.~(\ref{eq:3332}), (\ref{eq:polar-decoherence-vector}), and
(\ref{eq:g-damp}), change the properties of the $\Lambda$-system in a
nontrivial way. The susceptibility in the general case can be
calculated from Eq.~(\ref{eq:assymptotic-general-solution}) and
we find
\begin{align}
  \label{eq:general-chi}
  \chi
  =
  i\kappa
  \frac{
    \gamma/2\, \left(\eta_{{}_{-}}+\eta_{{}_{+}}\right)  |\Gamma_{+}|^2 \Gamma^*
    +
    \Big[
      \left(\gamma + \gamma_\mathrm{m} + 2 \eta_{{}_{-}}\right)
      \mathrm{Re} (\Gamma_{+}) \Gamma^*
      +
      \gamma (\eta_{{}_{-}}-\eta_{{}_{+}}) \Gamma_{+}^*
    \Big]
    |\Omega_{c}|^2
  }
  {
    \left(  \Gamma_{-}^*  \Gamma^* + |\Omega_{c}|^2\right)
    \Big[
    \gamma \eta_{{}_{+}} |\Gamma_{+}|^2
    +
    \left(\gamma + \gamma_\mathrm{m} - \eta_{{}_{-}} + 3 \eta_{{}_{+}} \right)
    \mathrm{Re} (\Gamma_{+})       |\Omega_{c}|^2
    \Big]
  },
\end{align}
where the parameters are defined in Eq.~(\ref{eq:parameters}). It should be noted that this 
expression contains all information of the detunings and phases of the optical fields, as well 
as the effects due to any combination of depolarization, dephasing, and $b \leftrightarrow c$ 
channels. Hence, the linear susceptibility of the probe field is obtained for any point in 
parameter space spanned by all of the above mentioned parameters, assuming only the validity 
of the Lindblad equation and the rotating wave approximation.

In the following we use Eq.~(\ref{eq:general-chi}) to study the susceptibility due to the 
different open system effects. The analytical expressions below are retrieved from
Eq.~(\ref{eq:general-chi}) as special cases by setting all channel rates  
except one to zero. We assume the non-zero channel rate $\eta_k$ to be much smaller than 
the decay rate, i.e., $\eta_k \ll \gamma$, and a Rabi frequency 
strong  enough to establish EIT, i.e., $\eta_k\gamma \ll |\Omega_{c}|^2$. Furthermore, we 
assume zero mean detuning $\Delta=0$, a balanced decay $\gamma_\mathrm{m}=0$, and that 
the two-photon detuning is not much larger than $|\Omega_{c}|$.

\begin{figure}[t]
  \centering
\includegraphics[width=0.95\textwidth]{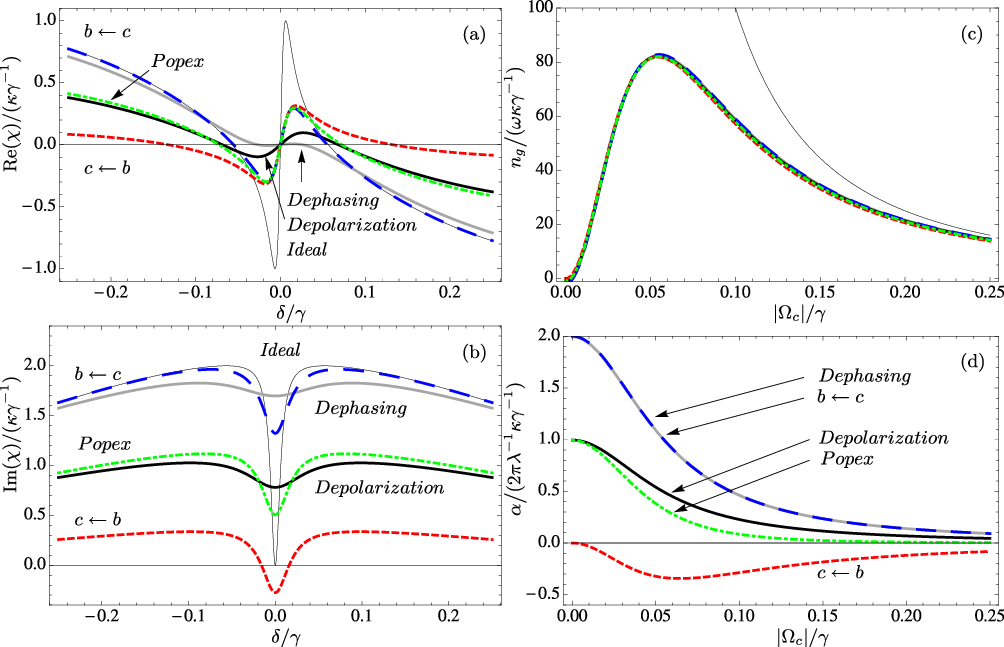}
\caption{(Color online) In (a) and (b) we show the real
and imaginary part of the susceptibility. In (c) and (d) the
slow-down factor (group index) $n_g$ and absorption $\alpha$ at two-photon resonance, are shown
as a function of the control field $|\Omega_c|$. The thin black line
shows the case of no open system interaction on $V_{bc}$. The thick
gray line shows the dephasing channel and the
thick black line shows the depolarization channel. The remaining lines
describe three versions of the amplitude damping channel: the
wide-dashed (blue) line shows the $b \leftarrow c$ damping channel, the
dashed (red) line shows the $c \leftarrow b$ damping channel, and the
dash-dotted (green) line shows the population exchange channel (popex), i.e., the incoherent sum of the two damping channels with equal rate. In all plots the proportionality factor
$\kappa$ defined below Eq.~(\ref{eq:susceptiblity-def}) is
scaled to 1. In (a) and (b) the parameter values are $|\Omega_c|=0.16$,
$\gamma=1$, $\Delta=0$, and all non-zero channel rates are set equal
to $0.1$. In (c) and (d) the channel rates are normalized so that
the slow-down factors are matched.}
\label{fig:susceptibility}
\end{figure}

In Figs.~\ref{fig:susceptibility}(a) and \ref{fig:susceptibility}(b) we show the susceptibility 
for different open system channels, using the same rate for all open system channels. Since the 
susceptibility is a complex function, the real and imaginary parts are plotted separately. Note 
that in the figures we use the exact expression in Eq.~(\ref{eq:general-chi}).

If we put all channel rates to zero except $\eta_z$ we obtain the susceptibility 
\begin{align}
  \label{eq:chi-dephase}
  \chi_{\mathrm{dephase}}
  =
	\kappa \frac{  2 \delta +2 i \eta_{z} }{|\Omega_{c}|^2 - \delta  (2 \delta + i \gamma )} .
\end{align}
This expression for the susceptibility is frequently encountered in the
literature, see, e.g., Ref.~\cite{Scully:QuantumOptics:1997}.

If we put $ \eta_{{}_{-}} = \eta_{{}_{+}} = \eta_{bc}\neq 0$ we find the susceptibility 
of a medium that effectively takes population from $| c \rangle$ to $ | b \rangle$. We have,
\begin{align}
  \label{eq:chi-damp_bc}
  \chi_{b \leftarrow c} =\kappa \frac{  2 \delta +i \eta_{bc}/2 }{|\Omega_{c}|^2 - \delta  (2 \delta + i \gamma )}.
\end{align}
This expression for the susceptibility is also frequently encountered in the literature, see, e.g., 
Ref. \cite{geabanacloche1995eit}. It is interesting to note the similarity of this expression with
Eq.~(\ref{eq:chi-dephase}), although the open system channels are quite different. Nevertheless, 
in Figs.~\ref{fig:susceptibility}(a) and \ref{fig:susceptibility}(b) it can be seen that the phase 
response of $\chi_{b \leftarrow c}$ is larger than for $\chi_{\mathrm{dephase}}$. This means that 
a medium affected by pure dephasing cannot slow down light as efficiently as compared to a medium 
affected by the $b \leftarrow c$ channel, assuming that the channel rates are equal.

We now consider the amplitude damping channel in the $c \leftarrow b$ direction. The expression 
of the susceptibility in this case reads
\begin{equation}
  \label{eq:chi-damp-cb}
    \chi_{c \leftarrow b}
  =
  \kappa \frac{ 2 \delta -i \eta_{cb}/2}{|\Omega_{c}|^2 - \delta  (2 \delta + i \gamma )}.
\end{equation}
This channel shows a different behavior compared to all other considered channels, it shows 
gain where the other channels show absorption. That is, a slowly propagating light pulse will be
amplified due to the interaction with the environment described by the $c \leftarrow b$ channel.

Let us now consider amplitude damping channels with equal rates in both directions in the 
subspace $V_{bc}$, corresponding to the non-zero parameters $\eta_{bc} = \eta_{cb} \equiv \eta_{pe}$. 
This open system channel may be important in experimental situations where, e.g., atoms collide 
and the populations of the states undergo sudden incoherent exchange. We denote this as the population 
exchange channel `popex'. We obtain,
\begin{equation}
  \label{eq:popex}
  \chi_{\mathrm{popex}}
  =
  	\kappa \frac{2 \delta + i \gamma  \eta_{pe} ^2/(8 |\Omega_{c}|^2)}{|\Omega_{c}|^2 - \delta  (2 \delta + i \gamma )}.
\end{equation}
As expected, the absorption profile is located in between the profiles of $\chi_{b \leftarrow c}$ 
and $\chi_{c \leftarrow b}$, as can be seen in Fig.~\ref{fig:susceptibility}(b). To first order, 
the absorption profiles of the different amplitude damping channels is shifted by the
parameter $\eta_{{}_{-}}$. However, the phase response, i.e.,
$\re{\chi_{b \leftarrow c}}$ is not affected by this shift, and it is
approximately the same for all considered amplitude damping channels.

The susceptibility associated with the isotropic depolarization
channel ($\eta \neq 0$) is given by
\begin{align}
  \label{eq:chi-depol}
  \chi_{\mathrm{depol}}
  =
  \kappa \frac{2 \delta + 2 i \eta/3}{|\Omega_{c}|^2 - \delta  (2 \delta + i \gamma )}.
\end{align}
In experiments, e.g., where one uses laser beams and hot atomic gases, this open system 
interaction is relevant as it corresponds to coherently prepared atoms leaving the laser 
beam, replaced with atoms in a completely mixed quantum state. The susceptibility is 
plotted in Fig.~(\ref{fig:susceptibility}) and it can be seen that the phase response 
$\re{\chi_{\mathrm{depol}}}$ is located between the corresponding curves for the 
dephasing and the amplitude damping channels. Furthermore, the absorption curve 
$\im{\chi_{\mathrm{depol}}}$ is close to the absorption profile of the depolarization 
channel, but significantly smaller than the corresponding curves for the dephasing 
and $b \leftarrow c$ channels. With respect to absorption and phase response, the 
depolarization channel is therefore not as harmful as the dephasing channel.

When there is no open system interaction on $V_{bc}$ we retrieve the
ideal susceptibility,
\begin{align}
  \label{eq:chi-ideal}
  \chi_{\mathrm{ideal}}
  =
  \kappa
  \frac{2\delta}{|\Omega_c|^2 - \delta ( 2 \delta + i\gamma)},
\end{align}
included in Fig.~(\ref{fig:susceptibility}) as a reference.

Often one is interested in achieving the slowest possible group velocity in a medium 
consisting of $\Lambda$-systems. 
The slow-down factor, or group index, is proportional to the derivative of the real part of the 
susceptibility $n_g \equiv \omega \partial_{\delta}\re{\chi} /2$ at angular frequency 
$\omega$. To compare the absorption for a given slow-down 
factor at $\delta = \Delta = 0$ we normalize the channel rates to give the same slow-down factor. Approximating $\partial_{\delta}
\chi$ for small channel rates in $V_{bc}$ gives
\begin{align}
  \label{eq:slow-down-factor-linear-expansion}
\partial_{\delta}\re{\chi}\big|_{\delta=\Delta=0} &
  \approx \kappa \frac{2 |\Omega_{c}|^2-\gamma  \left(\frac{8 \eta }{3}+\eta_{bc}+\eta_{cb}+
  2 \eta_{pe} + 4\eta_z \right)}{|\Omega_{c}|^4}
\end{align}
thus, the normalized rates are $\eta_{z} = \eta_{bc}/4 =
\eta_{cb}/4 = \eta_{pe}/2 = 2\eta/3 $.
Using these normalized rates, we plot the slow-down factor as a function of the
control field in Fig.~\ref{fig:susceptibility}(c) to verify that they are equal. 
The absorption coefficient is defined as $\alpha = (2\pi/\lambda) \im{\chi}$. In Fig.~\ref{fig:susceptibility}(d) we show the absorption on two-photon 
resonance as a function of the control field.
We see that the different open system
channels give different absorption profiles, even though the slow-down
factors are equal in this regime. Furthermore, it is interesting to
note that the dephasing and the $b \leftarrow c$ amplitude damping
channels, with normalized channel rates, have identical absorption
characteristics. We see that the isotropic depolarization
and population exchange channels are quite similar in
Fig.~\ref{fig:susceptibility}(d), but more essentially, they have a
lower absorption than the dephasing and the $b \leftarrow c$ amplitude
damping channels. As already mentioned, the $c \leftarrow b$ damping
is special in that it yields a gain where the other channels yield
absorption of the probe field.

\section{Conclusions}
\label{sec:conclusions}

We analyze a three-level $\Lambda$-system interacting with a Markovian
environment. The Lindblad master equation is reformulated using the
Gell-Mann matrices as a basis. We rewrite this in a
convenient vector formalism that is independent of the specific matrix
representation of the SU(3) generators. An evolution matrix
corresponding to this vector form is worked out in detail and the
dynamics is written as an affine-linear matrix differential equation.
The general solution of this equation can be given in terms of the
Jordan normal form. The Jordan blocks of the evolution matrix are
generically one-dimensional and the real part of the spectrum is negative. From
this follows that the asymptotic solution in all cases we are
interested in are independent of the initial states and given by the
inverse of the evolution matrix multiplied by a vector that is given
by the open system interaction.

The linear optical response of a quantum system is given by the
asymptotic steady state solution of the evolution equation. We find an explicit and closed analytical form of the susceptibility for general open system dynamics, including pure dephasing, depolarization, and three different forms of amplitude damping channels.

The open system effects introduce a non-trivial behavior of the linear
optical response of the $\Lambda$-system. For example, we show that the isotropic depolarization channel is less detrimental to the
induced transparency and the corresponding slow-down effect, than is
pure dephasing. Furthermore, an amplitude damping channel can lead to
reduced absorption and even gain for the probe field.

\section*{Acknowledgments}
\label{sec:acknowledgements}
M.E., E.S., and L.M.A. acknowledge the Swedish Research Council for
financial support. E.S. acknowledges support from the National Research Foundation and the Ministry of Education (Singapore).

\appendix
\section{}

In this Appendix, we provide a detailed derivation of the evolution
matrix $M$ and the vector $\bs{b}$ for Lindblad-type master equations
of a three-level system. Define the vector products
\begin{eqnarray}
\label{eq:wedge-and-star}
\bs{\alpha} \wedge \bs\beta & = & f_{rst} \mathbf{e}_r \alpha_s
\beta_t,
\nonumber \\
\bs{\alpha} \star \bs\beta & = & d_{rst} \mathbf{e}_r \alpha_s
\beta_t,
\end{eqnarray}
where $\bs\alpha=\alpha_q\bs{e}_q$ and $\bs\beta=\beta_q\bs{e}_q$ are
arbitrary complex-valued vectors, $\{ {\bf e}_i \}_{i=1}^8$ is a real
orthonormal basis of $\mathds{R}^8$, $f_{rst}$ and $d_{rst}$ are the
antisymmetric and symmetric structure constants of the SU(3) algebra~\cite{arvnid1997ngp},
and we have used Einstein's summation convention (repeated indices
are summed). The wedge product `$\wedge$' is a higher-dimensional vector
product analog to the cross-product in three-dimensional space. The
star product `$\star\,$' is a vector product that has no counterpart
in three dimensions.

Equipped with these tools we may now rewrite each term in the right-hand
side of Eq. (\ref{eq:masterequation-coherencevector}).
First, the Hamiltonian contribution reads
\begin{eqnarray}
\frac{1}{i} [\bs\omega\cdot \bs\lambda, \bs x \cdot \bs\lambda] =
2\left( \bs \omega \wedge \bs x \right) \cdot \bs \lambda .
\end{eqnarray}
Secondly, the middle term of Eq. (\ref{eq:masterequation-coherencevector})
is readily evaluated as
\begin{align}
  \label{eq:L(1)}
  \mathcal{L}(\mathds{1})
  =
\sum_{\mathrm{k}} 2i \left( \bs g_{\mathrm{k}} \wedge \bs
g_{\mathrm{k}}^{\ast} \right) \cdot \bs\lambda ,
\end{align}
where each term $\bs g_{\mathrm{k}} \wedge \bs g_{\mathrm{k}}^{\ast}$ is purely
imaginary, and hence $\mathcal{L}(\mathds{1})$ is real as required. Note that,
for Hermitian $\hat \gamma_{\mathrm{k}}$ the corresponding vector $\bs g_{\mathrm{k}}$
is real and thus $ \bs g_{\mathrm{k}} \wedge \bs g_{\mathrm{k}}^*=0$. By using the
symmetries of the structure constants the last term of
Eq.~(\ref{eq:masterequation-coherencevector}) takes the form
\begin{eqnarray}
  \label{eq:L(x.Lambda)}
  \mathcal{L}(\bs x \cdot \bs \lambda)
  & = &
  \sum_{\mathrm{k}}
     \Big\{
     \Big[
     \bs g_{\mathrm{k}}^* \wedge (\bs g_{\mathrm{k}} \wedge \bs x)
     +
     \bs g_{\mathrm{k}} \wedge ( \bs g_{\mathrm{k}}^* \wedge \bs x)
     \Big]
     \nonumber\\
     & & +\frac{i}{4}
     \Big[
     (\bs g_{\mathrm{k}} \wedge \bs x ) \star \bs  g_{\mathrm{k}}^*
     +
     \bs g_{\mathrm{k}} \star (\bs x  \wedge \bs  g_{\mathrm{k}}^*)
     -2 ( \bs g_{\mathrm{k}}^* \wedge \bs g_{\mathrm{k}} ) \star \bs  x
     +
     3 (\bs g_{\mathrm{k}} \star \bs x)  \wedge \bs  g_{\mathrm{k}}^*
     +
     3 \bs g_{\mathrm{k}} \wedge (\bs x  \star \bs  g_{\mathrm{k}}^*)
     \Big]
     \Big\} \cdot \bs \lambda .
\end{eqnarray}
By identifying terms, we arrive at the full Lindblad master equation
in coherence vector form
\begin{eqnarray}
  \label{eq:master-eq-vector-form}
  \dot{\bs x}
  &=&
   2\bs \omega \wedge \bs x +
  \sum_{\mathrm{k}}
     \Big\{
     \frac{2i}{3}\bs g_{\mathrm{k}} \wedge \bs g_{\mathrm{k}}^*
     +
     \Big[
     \bs g_{\mathrm{k}}^* \wedge (\bs g_{\mathrm{k}} \wedge \bs x)
     +
     \bs g_{\mathrm{k}} \wedge ( \bs g_{\mathrm{k}}^* \wedge \bs x)
     \Big]
     \nonumber\\
     & & +\frac{i}{4}
     \Big[
     (\bs g_{\mathrm{k}} \wedge \bs x ) \star \bs  g_{\mathrm{k}}^*
     +
     \bs g_{\mathrm{k}} \star (\bs x  \wedge \bs  g_{\mathrm{k}}^*)
     -2 ( \bs g_{\mathrm{k}}^* \wedge \bs g_{\mathrm{k}} ) \star \bs  x
     +
     3 (\bs g_{\mathrm{k}} \star \bs x)  \wedge \bs  g_{\mathrm{k}}^*
     +
     3 \bs g_{\mathrm{k}} \wedge (\bs x  \star \bs  g_{\mathrm{k}}^*)
     \Big]
     \Big\}.
\end{eqnarray}
It should be emphasized that this expression
is valid for any $N$-dimensional quantum system provided the SU($N$) structure
constants are being used in the definition of the vector products $\wedge$
and $\star$. Next, define
\begin{eqnarray}
	\label{eq:Mb}
  	M^{(0)} \bs x
	&\equiv&
	2 \bs \omega \wedge \bs x,
\nonumber
  \\
  G_\mathrm{k}^{(+)} \bs x
	&\equiv&
	\bs g_{\mathrm{k}}^* \wedge  (\bs g_{\mathrm{k}} \wedge \bs x)
	+
	\bs g_{\mathrm{k}} \wedge ( \bs g_{\mathrm{k}}^* \wedge \bs x ),
\nonumber
  \\
	G_{\mathrm{k}}^{(-)} \bs x
 	&\equiv&
 	\frac{i}{4}
  	\Big[
  	(\bs g_{\mathrm{k}} \wedge \bs x ) \star \bs  g_\mathrm{k}^*
  	+
  	\bs g_\mathrm{k} \star (\bs x  \wedge \bs  g_\mathrm{k}^*)
  	-2 ( \bs g_\mathrm{k}^* \wedge \bs g_\mathrm{k} ) \star \bs  x +
  	\nonumber\\
  	&&+
  	3 (\bs g_{\mathrm{k}} \star \bs x)  \wedge \bs  g_\mathrm{k}^*
  	+
  	3 \bs g_{\mathrm{k}} \wedge (\bs x  \star \bs  g_{\mathrm{k}}^*)
  	\Big],
\nonumber
   \\
    \label{eq:inhomogen-matrix_app}
	\bs b_\mathrm{k}
	&\equiv&
    \frac{2i}{3}\bs g_\mathrm{k} \wedge \bs g_\mathrm{k}^*.
\end{eqnarray}
We can put the equations of motion on the affine-linear matrix differential equation form in Eq.~(\ref{eq:11})
by combining Eqs.~(\ref{eq:wedge-and-star}) and (\ref{eq:Mb}). This yields
\begin{equation}
\label{eq:evolution-matrix}
M=M^{(0)}+\sum_{\mathrm{k}}M_{\mathrm{k}}=M^{(0)}+\sum_{\mathrm{k}}\Big(G^{(+)}_{\mathrm{k}} +
G^{(-)}_{\mathrm{k}} \Big)
\end{equation}
and vector
\begin{eqnarray}
\label{eq:vector-b}
\bs b=\sum_{\mathrm{k}} \bs b_\mathrm{k},
\end{eqnarray}
where $k$ represents different open system channels.
Here,
\begin{eqnarray}
  \label{eq:hamiltonian-matrix}
  M^{(0)}_{rt} &=& 2 f_{rst} \omega_s,
  \nonumber\\
  \label{eq:Matrix-B}
  G^{(+)}_{\mathrm{k},rt} &=& f_{rsm} f_{mvt} C_{\mathrm{k},sv}^{(+)},
  \nonumber\\
  \label{eq:Matrix-C}
      G^{(-)}_{\mathrm{k},rt} &=&
    \frac{i}{4}(
    d_{rms} f_{mvt} - d_{rmt} f_{msv} + 3 f_{rms} d_{mvt})
    C_{\mathrm{k},sv}^{(-)},
    \nonumber\\
    \label{eq:inhomogen-matrix}
    b_{\mathrm{k},r} &=& f_{rvs} C_{\mathrm{k},sv}^{(-)}
\end{eqnarray}
with
\begin{equation}
C_{\mathrm{k},sv}^{(\pm)}=(g_{\mathrm{k},s}^{\ast}
g_{\mathrm{k},v} \pm g_{\mathrm{k},s} g_{\mathrm{k},v}^{\ast}),\label{eq:C-matrix}
\end{equation}
being symmetric $(+)$ and antisymmetric $(-)$ in the indices $s$ and
$v$. Furthermore, while $C_{\mathrm{k},sv}^{(+)}$ is real-valued,
$C_{\mathrm{k},sv}^{(-)}$ is purely imaginary and thus vanishes for
real-valued ${\bs{g}}_{\mathrm{k}}$.  The antisymmetric matrix
$M^{(0)}$ corresponds to the Hamiltonian of the system. The matrix
$G_{\mathrm{k}}^{(+)}$ is real-valued and symmetric. The matrix $G_{\mathrm{k}}^{(-)}$
and  the vector $\bs b_\mathrm{k}$ vanish for real-valued ${\bs{g}}_{\mathrm{k}}$, as
they both are proportional to $C_{\mathrm{k}}^{(-)}$. $G_{\mathrm{k},rt}^{(-)}$ is
real-valued but has no obvious symmetry in the indices $r$ and $t$.


\begin{thebibliography}{33}
\bibitem[{foo()}]{footnote_aat}
\bibinfo{note}{Adding a term $g_0 \hat{\mathds{1}}$ to
  $\hat{\gamma}_{\mathrm{k}}$ yields the change $\mathcal{L}(\rho) \rightarrow
  \mathcal{L}(\rho) - i[{\mathrm{Im}} (g_0 \bs g_{\mathrm{k}}^{\ast}) \cdot
  \bs{\lambda},\rho]$, which corresponds to the effective change $H \rightarrow
  H + \hbar {\mathrm{Im}} (g_0 \bs g_{\mathrm{k}}^{\ast}) \cdot \bs{\lambda}$
  in the Hamiltonian.}
\end{thebibliography}

\end{document}